\documentclass[prc,twocolumn,showpacs]{revtex4}
\usepackage{amsmath,amsfonts}
\usepackage{epsfig}
\usepackage{graphicx}
\newcommand{\dr}{{\rm d}}
\newcommand{\al}{\alpha}
\newcommand{\be}{\beta}
\newcommand{\ga}{\gamma}
\newcommand{\om}{\omega}
\newcommand{\ep}{\epsilon}

\newcommand{\bea}{\begin{eqnarray}}
\newcommand{\beq}{\begin{equation}}
\newcommand{\eea}{\end{eqnarray}}
\newcommand{\eeq}{\end{equation}}
\def\diadots{\mathinner{\mkern1mu\raise8pt\vbox{\kern7pt\hbox{.}}\mkern2mu
\raise2pt\hbox{.}\mkern2mu\raise-4pt\hbox{.}\mkern1mu}}

\begin{document}
\title
{Time scales in nuclear giant resonances}
\author{W.\ D.\ Heiss$^{1}$, R.\ G.\ Nazmitdinov$^{2,3}$, 
F.\ D.\ Smit$^{4}$}
\affiliation{$^{1}$National Institute for Theoretical Physics, \\
Stellenbosch Institute for Advanced Study, \\
and Institute of Theoretical Physics,
University of Stellenbosch, 7602 Matieland, South Africa\\
$^{2}$Department de F{\'\i}sica,
Universitat de les Illes Balears, E-07122 Palma de Mallorca, Spain\\
$^{3}$ Bogoliubov Laboratory of Theoretical Physics,
Joint Institute for Nuclear Research, 141980 Dubna, Russia\\
$^{4}$iThemba LABS, PO Box 722, Somerset West 7129, South Africa}
\begin{abstract}
We propose a general approach to characterise fluctuations of measured cross 
sections of nuclear giant resonances. Simulated cross sections 
are obtained from a particular, yet representative
self-energy which contains all information about fragmentations.
Using a wavelet analysis, we demonstrate the extraction of
time scales of cascading decays into
configurations of different complexity of the resonance. 
We argue that the spreading widths of collective excitations in nuclei 
are determined  by the number of fragmentations as seen in the power
spectrum. An analytic treatment of the wavelet analysis
using a Fourier expansion of the cross section confirms this principle.
A simple rule for the relative life times of states associated with hierarchies 
of different complexity is given.
\end{abstract}
\maketitle

Nuclear Giant Resonances (GR) have been the subject of numerous investigations over 
several decades \cite{dstadta}. Some of the basic features such as centroids and
collectivity (in terms of the sum rules) are reasonably well understood within 
microscopic models \cite{BM2,micro}. However, the question how a collective mode 
like the GR disseminates its energy is one of the central issues in nuclear structure 
physics. 

According to accepted wisdom, GRs are essentially excited by an
external field being a one-body interaction. 
It is natural to describe these states as collective 1p-1h states. 
Once excited, the GR disseminates its energy via direct particle emission and by coupling
to more complicated  configurations (2p-2h, 3p-3h, etc). The former mechanism gives
rise to an escape width, while the latter yields spreading widths ($\Gamma^{\downarrow}$).
An understanding of lifetime characteristics associated with 
the cascade of couplings  and scales of fragmentations 
arising from this coupling (cf \cite{zel,sok,aiba,lar}) remains a challenge. 
A recent high-resolution experiment of the Isoscalar Giant Quadrupole Resonance 
(QR) \cite{dstadtc,dstadtd,dstadte}  provides new insight for this problem.

It has been shown by Shevchenko {\it et al.} \cite{dstadtc} 
that the fine structure of the QR observed in $(p,p')$ experiments 
is largely probe independent. Furthermore, a study of the fine structure using 
wavelet analysis \cite{w1,w2,w3} reveals energy scales \cite{dstadtd,dstadte} 
in the widths of the fine structure displaying a seemingly schematic pattern, 
as can be seen in Fig.2. This pattern varies with the structure of the nucleus 
being studied. While the physical meaning of the results of such an analysis is 
still being debated, we try here to offer a general explanation. However, we do 
not embark on a specific microscopic analysis, but rather make use of general and 
well-established techniques of many-body theory. 
Gross effects due to nuclear deformation 
and coupling to the continuum \cite{sok} are not discussed;
we rather focus on the decay of the QR into configurations
of various complexity.

To proceed we use the Green's function
approach. A central role is played by the self-energy whose finer
structure is imparted upon the Green's function via the solution
of Dyson's equation which reads \cite{mbody}

\beq
G_{\al,\be}(\om)=((G^0_{\al,\be}(\om))^{-1}-\Sigma
_{\al,\be}(\om))^{-1}
\label{green}
\eeq
where we assume $G^0(\om )={\delta_{\al,\be}}/{(\om -\ep)}$
to be diagonal in the basis $\al ,\be ,\ldots $ while the complicated
pole structure of $G(\om )$ is generated by that of the self-energy
$\Sigma _{\al,\be}(\om)$. The pole structure of $G$ carries over to
the scattering matrix given by
\beq
T_{\al,\be}(\om)=\Sigma _{\al,\be}(\om)+
\Sigma _{\al,\be'}(\om)G_{\be',\al'}(\om)\Sigma _{\al',\be}(\om)
\label{scat}
\eeq
from which a cross section $\sim |T_{\al,\al}(\om)|^2$ is obtained.

Within the excitation energy range of the QR the 
nucleus has a high density of complicated states of several tens of thousands 
per MeV and even more for heavy nuclei. These many states appear in the self-energy 
as poles  in the complex energy plane close to the real axis. The small
widths imply they are long-lived states and traditionally classed as compound
states. The simpler intermediate structure of the excitation is
expressed by the substantial fluctuations of the corresponding
residues associated with the poles of the self-energy
$\Sigma(\om )$ \cite{hahe}. In other words, while the individual pole positions 
of $\Sigma(\om )$ are virtually unstructured \cite{random}, it is the variation 
of the corresponding residues that bears all the information about intermediate 
structure. Note that our approach differs from a traditional
microscopic calculation in that we start from the outset from a random
distribution of pole terms representing compound states. Traditional
microscopic approaches cannot do justice to such structure  \cite{ves}
and usually suffer from necessary truncation of configuration space
and, associated with it, from possible inconsistencies and spurious states.

We assume that the QR being a collective 1p-1h state decays via
a cascade progressing through (2p-2h)-, (3p-3h)-configurations and so forth to 
the eventual compound states. In turn, each of the intermediate states
(including the initial QR) can either decay directly to the ground
state or via some more complicated intermediate state. Below we will show that 
it is this mixture that is seen in the cross section and extracted by wavelet analysis,
and it is the variety and cascading complexity of states that invokes
the structure of the residues of the poles of the self-energy.
Of importance to note is that the number of states available within the
energy domain of the QR increases with its complexity: for example, six
(2p-2h)-states, eleven (3p-3h)-states, down to several thousand
compound states (the numbers six or eleven should be taken as examples
without claim for quantitative correctness). Moreover, the
corresponding life times are expected to increase in line with their 
increasing complexity, which is in accordance with their decreasing 
spreading widths (below we come back to this particular aspect of scaling).
\begin{figure}[t]
\epsfig{file=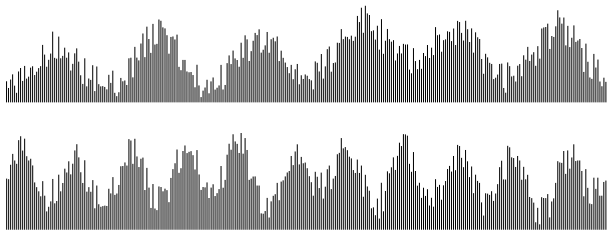,height=3.0cm,width=8.0cm,clip=,angle=0}
\caption{Residues of the self-energy: for 6 (top) and 11 (bottom)
intermediate states. The randomisation is clearly discernible.}
\label{tria}
\end{figure}

As a typical case study we investigate here a wavelet
analysis of a simulated cross section
that results from a particular input for the self-energy. Since
arbitrary units are used, we concentrate on the energy interval [0,1]
and use for the pole position $\epsilon=0.5- i 0.5$
of the single pole of $G_0$ (\ref{green}).
The number of compound states is assumed to be 300;
this is of course much less than the experimental level density in the region of
a QR for a medium or heavy nucleus, but it suffices for our demonstration.
The real parts of the pole positions are assumed to be randomly
distributed with a uniform distribution of the mean distance 1/300; the
imaginary parts are randomly distributed in the interval
[0.004,0.007].

For the sake of illustration we consider four sets of residues
\beq
r_k=\sum _{i=1}^4h_{i,k},\quad 
 h_{i,k}= s \sum _{j=1}^{f_i}\frac{\ga _i^2}{(\frac{k}{300}-j\cdot p_i)^2+\ga_i^2}
\label{res}
\eeq
with an overall strength $s=10^{-5}$. This order of magnitude is based
on the mean value of the widths of the compound states
being about $10^{-4}$ to $10^{-5}$ times smaller than the
$\Gamma^{\downarrow} \;(\ga _i)$. 
With these residues the self-energy reads
\beq \Sigma (\om)=\sum _{k=1}^{300} \frac{r_k}{\om-\om_{k}}
\label{sepole}
\eeq
 The poles at the complex positions
$\om _{k}$ occur in the lower $\om $-plane with $\om $ being the energy
variable; the other symbols are explained in the text.
If only $i=1$ was to occur with $f_1=6$, a typical pattern of
the residues $r_k=h_{1,k}$ would be illustrated by the top of
Fig.1; similarly for $f_2=11$ by the bottom.
The inclusion of further terms would simply add additional peaks to the pattern.
In the case presented below we have
chosen $f_2=11,f_3=17$ and $f_4=29$ totalling to 6+11+17+29 additional
peaks (not easily visualised, but beautifully discernible in the final analysis).
We stress again that the four values $f_i$ were chosen for
demonstration, more than four or other values can be used just as well.

These arbitrary numbers used in the example chosen describe particular
fragmentations of the QR into altogether 6,11,17 and
29 states of increasing complexity. The widths $\ga_i$ giving rise to
the Lorentzian shape of the residues are in reality determined by the
product of the density of the compound states and the coupling of the
$i$-th  group to the compound states. The widths are
the spreading widths of the respective states considered \cite{hahe}.
As the complexity
increases with label $i$ we shall assume $\ga_1>\ga_2>\ga_3>\ga_4$.
In the simulation we endow each $\ga_i$ with a random fluctuation with
mean value $\ga_i/4$. As stated above we refrain from specifying a
microscopic structure causing the residue pattern assumed for the self-energy;
below it becomes clear that guidance comes from experiment.

\begin{figure}[ht]
\epsfig{file=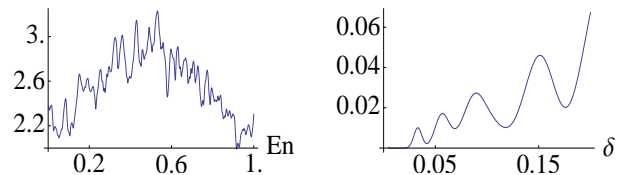,height=2.2cm,width=8.0cm,clip=,angle=0}
\caption{Simulated cross section (left) and power spectrum (right).
Units are arbitrary. The
abscissa of the cross section is the unit energy interval, the energy
values $\delta $ on the abscissa of the power spectrum refer to the wavelet parameter
using the same energy units.}
\label{crosss}
\end{figure}

We also assume that each set ${f_i}$ uniformly distributed
over the whole energy interval. This is similar in spirit to the assumption used in
the local scaling dimension approach \cite{aiba}. The positions $p_i$ in Eq.(\ref{res})
are set to be $\sim 1/f_i$ which spreads the actual $j\cdot p_i$ positions equidistantly 
over the whole interval  with $j$ running from 1 to $f_i$; however, we endow them 
with a small random fluctuation with mean value $p_i/8$. Note that
the random fluctuation of widths and positions generate a mild degree
of asymmetry in the energy interval [0,1], resulting
in slightly different patterns in the intervals [0,0.5] and [0.5,1].
The near equality of the positions, that is - apart from slight random
fluctuations - the regular pattern of the various fragments as
illustrated in Fig.1, is basically dictated by
experimental findings: {\sl if there is no near regular pattern there will
be no discernible structure in the power spectrum of the wavelet
analysis}.  However, we shall
return below to the case where such regular pattern may occur only in a
smaller portion of the interval.

The first obvious choice for the widths assumes simply $\gamma_i=1/(2 f_i)$
yielding the simulated cross section shown in Fig.2 
(below a precise analytic expression confirming the $1/(2f_i)$-law is given). 
A variation of such choice is rather significant, we shall return to this aspect 
in detail. 

\begin{figure}[t]
\epsfig{file=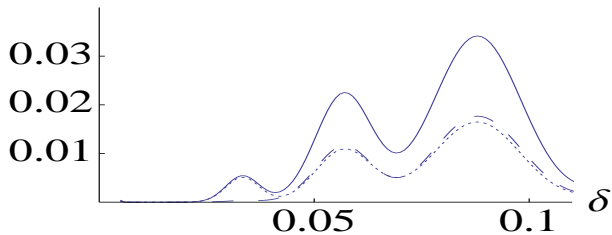,height=3.0cm,width=8.0cm,clip=,angle=0}
\caption{Power spectrum for a particular asymmetric situation discussed in text.
The dotted curve originates from a scan of the interval
[0,0.5], the dashed curve from [0.5,1] and the solid curve from the
total interval. Note that the most left peak is virtually absent
in the dashed curve while fully present in the dotted curve. 
Units as in Fig.2.}
\label{asym}
\end{figure}

The analysis using a Morlet-type mother wavelet (a contour plot of the wavelet 
analysis is illustrated in Fig.5)
\beq
\frac{1}{\sqrt{\delta }}\,\cos \frac{k (\om -\om_0)}{\delta }\, \exp
-\frac{(\om -\om_0)^2}{\delta ^2}
\label{mor}
\eeq
gives the power spectrum shown
in Fig.2; if not indicated otherwise we use the value $k=6$
for the wave number of the mother wavelet.
There is in fact a $k$-dependence of the
positions of the maxima of the power spectrum, it is given in analytic
terms below.

On the right part of Fig.2 we clearly discern the four maxima that are produced by the
four different values $f_i$ of the number of fragmentations. In fact, the
fragmentation into $f_1=6$ produces (for $k=6$) the maximum roughly at
$\delta^{\rm max}_1=1/f_1=1/6$;
similarly, the other three maxima occur at $\delta^{\rm max}_i=1/f_i,\,i=2,3,4$. This is one of our
major findings:
\par \noindent
{\sl the maxima of the power spectrum occur at 
$$\delta ^{\rm max}_i\approx k/(2\pi )\cdot I/f_i$$ with $I$ being
the interval of the whole range of the QR considered and $f_i$ the number of
fragmentations.} The factor $k/(2\pi )$ originates from the analytic
expression given in (7) below.

The asymmetry found in some experimental data can obviously be accounted
for by our analysis. We refer to cases where the
analysis yields a pattern in the first half of the whole resonance being
different from that in the second half, or in principle for any
subdivision of the whole resonance. For illustration, we 
take $f_4=14$ while leaving all other parameters unchanged. In this way the 
total of 29 maxima of the residues
$r_{f_4}$ are confined to only 14 within the left half of the interval. The
effects are clearly seen in Fig.3. Note that the positions
of the maxima still remain unchanged. This type of asymmetry is
clearly discernible in Fig.9 of Ref.\onlinecite{dstadte}: from the
two-dimensional wavelet transform the wavelet power would give
a similarly different pattern when taken at different portions of the
whole interval.

\begin{figure}[b]
\epsfig{file=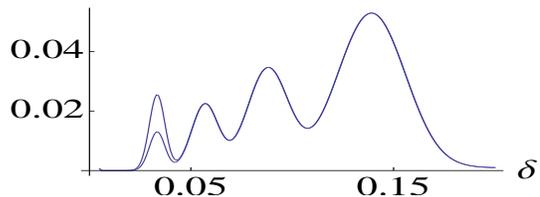,height=2.5cm,width=7.0cm,clip=,angle=0}
\caption{Power spectrum: dependence of height at the maximum on spreading width.
The curve with the lower value of the most left maximum
is identical to the one on the r.h.s. in Fig.2., while the higher peak
is due to a decrease of its spreading width or an increase of its life time. Units as in Fig.2.}
\label{width}
\end{figure}
The folding (integration) of the cross section with the Morlet
wavelet has to be done numerically. In order to obtain an analytic
expression relating the number of fragmentations $f_i$ to the
positions of the maxima of the power spectrum, we consider an
expansion of a cross section into a Fourier series 
\beq
\frac{\dr \sigma(\omega)}{\dr \Omega}=\sum _m c_m \sin (m \pi \om/I) +c_0
\label{cs}
\eeq
with the bulk term
$$c_0=\int  \frac{\dr \sigma(\omega)}{\dr \Omega} \, \dr \omega $$ (further terms
with $\cos (m \pi \om/I)$ are immaterial for the discussion). 
An intermediate structure  manifests itself, if a few terms in (\ref{cs}) are 
appreciably stronger than the others. In Fig.2 the terms with
$c_{12}\approx c_{22}\approx c_{34}\approx c_{58}$ are dominant; of course, terms 
for different $m$-values also occur but are smaller by roughly an order of magnitude 
or more (here our analysis does not focus on $m\le 4$: while giving larger
contributions such values would correspond to $\delta \ge 0.5$ and represent gross 
and bulk structure). Performing analytically the wavelet-transform of each term in 
(\ref{cs}) ({\sl Mathematica} gives a closed expression for the integral from which 
the formula below can be extracted), one obtains an analytic evaluation of the 
positions and heights of the maxima of the power spectrum. For each 
$\sin (m \pi x)$-term the positions of the local maxima in the power spectrum turn 
out  to be
\beq
{\rm Max}_m=\frac{k+\sqrt{2+k^2}}{2 m \pi }I.
\label{max}
\eeq
For $k=6$ (and the unity interval $I$) this yields 0.16, 0.088, 0.057 and 0.033 for $m=12,22,34$
and 58, respectively as verified in Fig.2. Note that a different choice of $k$ moves
the positions of the local maxima, yet the $\sim 1/m$ law prevails.
The expression (\ref{max}) provides an obvious tool to
be used to ascertain the number of fragmentations
when the maxima are determined from an analysis of experimental data.
Clearly, the number $f_m$ of fragmentations
introduced above is related to the value $m$ in (\ref{cs}) by $m=2f_m$.

Furthermore, an increased value of $k$ can resolve a peak in the
power spectrum that is caused by two near values of $f_i$.
In fact, the distance between adjacent maxima (say $m=17$ and $m=18$)
roughly doubles when $k$ is doubled.

While - for fixed $k$ - the $1/f_i$ dependence of the maxima 
of the power spectrum is an important finding, even more significant is the result 
that the values at the maxima (the heights) also obey
the same $1/f_i$-law {\sl if the corresponding Fourier coefficients
  are about equal}. 
Indeed, a straight line can be drawn through the maxima in Fig.2 as 
the four values $c_m,\,m=12,22,34,58$ are about equal. We recall that, for example,
$\sin (12 \pi x)$ generates $f_k=6$ peaks of a width $\gamma _k=1/(2f_k)$ in 
the energy (unit) interval for the cross section. 
This can be exploited in a realistic analysis:
a deviation from this straight-line-rule signals effectively a deviation from the spreading width
being assumed to be $1/(2 f_i)$. This is illustrated in Fig.4 where the spreading width
$1/(2f_4)$ has been decreased to $1/(2.8f_4)$. As a result, the value of the first peak
becomes enhanced. Since the spreading width is related to the life time of the states, 
we conclude:
{\sl the life times are proportional to $f_i$ if the heights of the maxima lie on 
a straight line; an increased (decreased) height signals an even longer (shorter)
life time}.

In this context we note that the {\it number} of peaks and troughs in Fig.5 on 
the horizontal lines matches exactly the values of the $f_i$: six on the top, 
further down eleven, then seventeen and twenty nine on the bottom. The actual 
values of these peaks and troughs determine the heights of the bumps in the 
power spectrum, that is the information about the life times
of the respective fragmented states. A similar wavelet transform
obtained from experimental data is presented in Fig.8 (and 9) in 
Ref.\onlinecite{dstadte}; note that our schematic 'in vitro' illustration is of
course much more symmetric.

While in experiments the chaotic nature of the nucleus usually shows at higher excitation
energies \cite{random}, the pertinent structure revealed in the analysis may come as a surprise.
We are of course familiar with order in the nuclear many body system
as shown in shell effects and simple collective
states. The fragmentations of the QR may be due to a different quality: it
could be a manifestation of {\em self-organising structures} \cite{bak,nil,sor}.
Indeed, the life-time of increasingly complex configurations of the QR is increasing toward 
the compound states and the ground state. There is no general accepted definition of 
conditions under which the self-organising structures are expected to arise. 
We may speculate that in the case considered here, 
once the nuclear QR state is created, it  is driven 
to an unstable hierarchy of configurations (metastable states) by
quantum selection rules which connect these different complex configurations
due to internal mixing. This problem needs of course a dedicated study on its own and 
is beyond the scope of the present paper.

\begin{figure}[t]
\epsfig{file=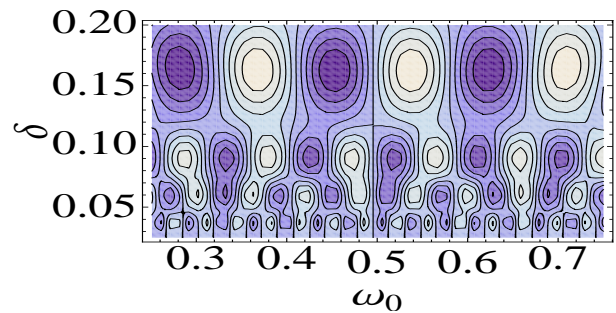,height=4.0cm,width=8.0cm,clip=,angle=0}
\caption{Wavelet contour plot of cross section
shown in Fig.2. The symbols $\delta $ and $\om _0$ refer to the Morlet wavelet
parameters used in (\ref{mor}). 
The (positive) maxima are in light shading and the (negative)
minima in dark. For the top pattern the contours range from 0.4 to $-0.4$.}
\label{wavel}
\end{figure}

We summarise the major points of our findings:
(i) the position of the peaks in the power spectrum indicate the
number of fragmentations of a particular intermediate state; the
more complex states lie to the left of the simpler states (see Eq.(\ref{max}));
(ii) the resolution of poorly resolved peaks
can be improved by a higher value of $k$; (iii) the values (heights)
at the peaks are related to the spreading widths, implying knowledge about the life
times: if they lie on a straight line, the life times are proportional
to the number of fragmentations, if they lie above (below) the
straight line the corresponding life times are longer (shorter).
Finally, we mention that a pronounced gross structure of the
experimental cross section as found in lighter nuclei, would have no
effect upon our findings. In fact, such gross structure had to occur
at the far right end (values of $\delta $ appreciably larger than
those used in the literature) of the power spectrum.

\section{Acknowledgement}
WDH is thankful for the hospitality which he received
from the Nuclear Theory Section of the Bogoliubov Laboratory, JINR during his
visit to Dubna. The authors gratefully acknowledge enlightening
discussions with J. Carter, R. Fearick and P. von Neumann-Cosel.
This work is partly supported by JINR-SA Agreement on scientific collaboration,
by Grant No. FIS2008-00781/FIS (Spain) and
RFBR Grants No. 08-02-00118 (Russia).

\end{document}